\documentclass[10pt,a4paper,twoside]{article}

\usepackage{indentfirst}			
\usepackage{graphicx}				
\usepackage{amsgen,amsfonts,amssymb,amsbsy}	

\newenvironment{resum}{\begin{quote}\small}{\end{quote}}

\newcommand{\bfsf}[1]{\textsf{\textbf{#1}}}

\begin{document}

\thispagestyle{plain}		

\begin{center}


{\LARGE\bfsf{On the Penrose Inequality}}

\bigskip


\textbf{Edward Malec}$^1$, 
\textbf{Marc Mars}$^2$ and
\textbf{Walter Simon}$^3$


$^1$\textsl{ Uniwersytet Jagiello\'nski, Krak\'ow, Poland} \\
$^2$\textsl{Universidad de Salamanca, Spain}\\
$^3$\textsl{Universit\"at Wien, Austria}

\end{center}

\medskip


\begin{resum}

We summarize results on the Penrose inequality bounding the ADM-mass or the Bondi mass
in terms of the area of an outermost apparent horizon for asymptotically flat initial data 
of Einstein's equations. We first recall the proof, due to Geroch and to Jang and Wald, 
of monotonicity of the Geroch-Hawking mass under a smooth inverse mean curvature flow
for data with non-negative Ricci scalar, which leads to a Penrose inequality if the apparent 
horizon is a minimal surface.
We then sketch a proof of the Penrose inequality  of Malec, Mars and Simon which holds for 
 general horizons and for data satisfying the dominant energy condition, but imposes (in 
addition to smooth inverse mean curvature flow) suitable restrictions on the data
on a spacelike surface. These conditions can, however, at least locally be fulfilled by a 
suitable choice of the initial surface in a given spacetime. Remarkably, they
are  also (formally) identical to ones employed earlier by Hayward in order to define a 2+1 
foliation on null surfaces, with respect to which the Hawking mass is again monotonic.
\end{resum}

\bigskip


In an attempt for getting evidence on the "cosmic censorship conjecture"
in asymptotically flat spactimes, Penrose considered collapsing shells of particles with
zero rest mass and gave a heuristic argument in favour of the inequality

\begin{equation}
\label{pi}
M_{ADM} \ge \sqrt{\frac{A}{16\pi}}
\end{equation}
between the  ADM mass $M_{ADM}$ measured at spatial infinity and the area $A$ of an 
apparent horizon \cite{RP}. Another conjecture with the Bondi mass $M_{B}$ measured at null 
infinity instead of $M_{ADM}$ can also be formulated.

To establish this "Penrose inequality" (PI) for a selected apparent horizon in a given spacetime 
$({\cal M},{}^4g_{\mu\nu})$ we have to  connect this horizon with infinity by a 
hypersurface $\cal N$. To describe this procedure in some detail we recall some facts 
about initial data sets $({\cal N},g_{ij},k_{ij})$, with apparent horizons as boundaries.

The induced metric $g_{ij}$ on $\cal N$ and the second fundamental form $k_{ij}$ are 
required to satisfy the constraints
\begin{eqnarray}
\label{con1}
D_i \left( k^i_l - k \delta^i_l \right) & = & - 8 \pi j_l, \\
\label{con2}
R - k_{ij}k^{ij} + k^2 & = & 16\pi\rho.
\end{eqnarray}
where $D_{i}$ is the covariant derivative and $R$ the Ricci scalar on 
${\cal N}$, $k = g^{ij}k_{ij}$, and $\rho$ and $j_{i}$ are the energy density and the matter 
current, respectively. We take the data to be asymptotically flat and to satisfy the dominant 
energy condition (which implies that $\rho \ge |j|$).

For  any 2-surface ${\cal S}$, we denote the tangents to the outgoing future and past null 
geodesics orthogonal to ${\cal S}$ by $\vec L^{\pm}$, with inner product $\varphi =
\langle \vec L^{+}, \vec L^{-} \rangle$. 
We introduce the "expansions" of $L^{\pm}$ by
\begin{equation}
\label{exp}
\Theta^{\pm} = (g^{\mu\nu} - 2 \varphi^{-1} L^{+(\mu} L^{- \nu)})\nabla_{(\mu}L_{\nu)}^{\pm}
\end{equation}
which clearly depend on the normalization of the null vectors.

A ``future apparent horizon'' ${\cal H}$  (called a ``horizon'' from now on)
is a 2-surface defined by the property that all outgoing future directed null
geodesics in ${\cal M}$ orthogonal to ${\cal H}$ have vanishing divergence,
i.e. $\Theta^{+} = 0$, and that the divergence of the outgoing past null
geodesics is non-negative, i.e. $\Theta^{-} \ge 0$. 
(The term "outgoing" is defined with respect to the asymptotically flat end.
Past apparent horizons are defined analogously). If $\Theta^{+} = \Theta^{-} = 0$ 
on ${\cal H}$, the horizon is an extremal surface (in $({\cal M},g_{\mu\nu})$ as well as
in any submanifold ${\cal N}$) and the outermost extremal surface in an asymptotically flat space 
must be minimal.

It is now useful to distinguish the following three cases.

\begin{description}
\item[The "Riemannian case":]
This common (but somewhat unhappy) terminology denotes the  case where ${\cal N}$ is bounded by a minimal 
surface (instead of a general trapped one) and where (instead of assuming the constraints) the Ricci scalar 
on ${\cal N}$ is required to be non-negative. Such 3-manifolds arise as spacelike slices of a spacetime with a 
horizon only in very special situations such as in the time-symmetric case.
\item[The general spacelike case:]
Here one joins a chosen apparent horizon via a spacelike surface either with spatial infinity or with a cross 
section of null infinity.
\item[The null case:]
One joins a chosen (future) apparent horizon with a cross section of (past) null infinity by a null surface.
\end{description}

Mathematical proofs of the PI with "realistic" smoothness assumptions have so far only 
been obtained in the Riemannian case. The strongest result to date is due to Bray \cite{HB} 
who proves (\ref{pi}), admitting also disconnected minimal surfaces with (total) area $A$.
However, we rather wish to expose here in some detail other (older) approaches which produce 
weaker results in the Riemannian case but are more promising, and also have produced some 
partial results, in the other cases. These approaches are based on defining {\it flows} and 
corresponding {\it 2 + 1 decompositions} of the chosen hypersurfaces. In particular the 
following steps can be identified in this strategy:

\begin{enumerate}

\item Choose a suitable 3-surface ${\cal N}$ joining the apparent horizon with infinity.

\item Choose (or if possible, \textit{show existence !}) of a (\textit{weak}) flow on
${\cal N}$ starting at the horizon and going to infinity. Consider the corresponding 
foliation $S(r)$ by level surfaces of the flow (\textit{if it exists !}).

\item Define a "quasilocal mass"  $M_{QL}(r)$ as a surface integral, in particular over the level sets of the flow.

\item Show that $\lim_{r \rightarrow \infty} M_{QL} \le M_{ADM}$ or that 
$\lim_{r \rightarrow \infty} M_{QL} \le M_{B}$ (where the "end" of ${\cal N}$ is at $r \rightarrow \infty)$.

\item Show that $M_{QL}(r_{\cal H}) \ge \sqrt{A/16 \pi}$ (where $r_{\cal H}$ denotes the horizon).
 
\item Show monotonicity of the mass along the flow, i.e. $(d/dr) M_{QL}(r) \ge 0$.

\end{enumerate}

It is obvious that the PI, equ. (\ref{pi}), follows if these steps can be carried out. 
In practice, it is useful to restrict oneself first to smooth flows, and to try to generalize afterwards.
We are now going to describe the 3 cases identified before.\\ \\
{\bf The Riemannian case.}

We first assume that there is a smooth ``inverse mean curvature flow'' (IMCF) starting from a minimal
surface. This means that one can write the metric $g_{ij}$ as

\begin{equation}
\label{met}
ds^2 = \phi^2 dr^2 + q_{AB}dx^A dx^B
\end{equation}
(where $A,B = 2,3$), with smooth fields $\phi$ (the "lapse") and $q_{AB}$, and with $p \phi =
g(r)$ where $p$ is the mean curvature of the surfaces $r= \mbox{const.}$ and
$g(r)$ is a smooth function. In other words, $p \phi$ is constant on the
level sets of the flow. Note that $\phi$ diverges at the minimal surface, but the
integral curves of the flow can be extended. Assuming also that the level sets  
have spherical topology, Geroch showed that the mass functional
\begin{equation}
\label{mg}
M_{G}({\cal S}) = \frac{\sqrt{A}}{64\pi^{\frac{3}{2}}}
\left (16\pi - \int_{\cal S} p^2 dS \right)
\end{equation}
(taken over level sets of the IMCF) is monotonic \cite{RG}, while Jang and Wald observed that this leads 
to the proof of the PI along the lines sketched above \cite{JW}. Later  Huisken and Ilmanen succeeded in 
removing the restrictive and "unphysical" assumption of smoothness of the flow \cite{HI}. These authors 
 established the existence of weak solutions of the degenerate elliptic equation for the level sets 
of the IMCF. The resulting generalized flow appears to "jump" at (countably) many places to  so-called 
minimizing hulls. Most importantly, the Geroch mass increases at each jump, and so monotonicity is preserved. 
In this setup it is also possible to handle disconnected horizons but the
presence of jumps prevents obtaining the PI with the total area of the
horizon.  Unlike Bray's result, the PI proved by this method  only
 contains the area of any connected component selected to start the 
IMCF. (In particular, the largest component can be chosen for this purpose).\\ \\
{\bf The general spacelike case.}

Let $({\cal N},g_{ij},k_{ij})$ be a data set with ${\cal N}$ spacelike, and assume as before that there is a 
smooth IMCF with level sets of spherical topology. Let  $\vec M = d/dr$, or equivalently $M^{i}D_{i}r = 1$, 
denote the tangent vector field to the flow. We normalize the null vectors $\vec L^{\pm}$ emanating 
orthogonally from the level sets  by requiring that 
\begin{equation}
\label{lsum}
\vec L^{+} + \vec L^{-} = \sqrt{2} \vec M.
\end{equation}

Here we describe (and rewrite sligthly) a result of Malec, Mars and Simon \cite{MMS} who show monotonicity of the 
Hawking quasilocal mass \cite{SH}
\begin{equation}
\label{mh}
M_{H}({\cal S}) = \frac{\sqrt{A}}{64\pi^{\frac{3}{2}}}
\left (16\pi - \int_{\cal S} \varphi^{-1} \Theta^{+}\Theta^{-} dS \right)
\end{equation}

(defined in terms of the expansions (\ref{exp})) under the following restrictions of the data 
(in addition to smooth IMCF):
\begin{eqnarray}
\label{thpos}
\Theta^{-} & \ge & 0~~ \mbox{on the level sets of the IMCF on}~{\cal N}, \\
\label{thquot}
\frac{\Theta^{+}}{\Theta^{-}} & = & f(r)~\mbox{for some smooth function}~f(r). 
\end{eqnarray}

Equ. (\ref{thquot}) means that $\Theta^{+}/\Theta^{-}$ is constant on the level sets of the IMCF.
Note
that $M_{H}$ does not depend on the normalization of the null vectors $L^{\pm}$, whereas condition (\ref{thquot}) 
is to be understood in terms of the normalization introduced above.
We next remark that, again using this normalization, 
we have  $\sqrt{2} p \phi = \Theta^{+} + \Theta^{-}$. Hence the condition of IMCF, together with (\ref{thquot}), 
is equivalent to requiring the existence of a foliation on which both expansions  $ \Theta^{+} $ and $\Theta^{-}$ 
are constant. We will refer to this observation when describing the null case below.
Moreover, equ.(\ref{thquot}) together with IMCF is also equivalent with 3.(a) in the theorem of \cite{MMS}
(where a different normalization was used).

We finally note that (\ref{thquot}) can locally be fulfilled by a suitable choice of the hypersurface in a given 
spacetime. In fact, consider an arbitrary (smooth) function $F(r)$  with $F(0)=-1$ and $|F(r)| < 1$ for $r > 0$. 
Rescaling the null vectors $\vec L^{-}$ and $\vec L^{+}$ emanating from ${\cal S}$ like 
$\vec L^{+} \rightarrow \lambda \vec L^{+}$ and $\vec L^{-} \rightarrow \lambda^{-1} \vec L^{-}$, with 
$\lambda^2 = \Theta^{-} (\Theta^{+})^{-1}(1 + F(r))(1 - F(r))^{-1}$  the relation (\ref{lsum}) then determines the 
tangent to the hypersurface on which the IMCF satisfies (\ref{thpos}) as well as (\ref{thquot}). In view of the large 
freedom for choosing $F(r)$ it is also plausible that a suitable choice would make the surface ${\cal N}$ reach 
spacelike infinity.\\ \\

{\bf The null case.}

Ludvigsen and Vickers have shown that the "Nester-Witten" quasilocal mass is monotonic along a suitable flow on a 
surface generated by null geodesics emanating orthogonally from the apparent horizon \cite{LV}. More recently 
Bergqvist has substantially simplified the proof of this result by removing all the spinors from the arguments \cite{GB}.
The problems with applying these results to a proof of the PI are, firstly, that the null surfaces constructed in 
this way do not exist globally in general and secondly, that the quasilocal masses are not known to
be bounded by the Bondi mass at infinity in the sense of condition 4 above. 

In contrast, Hayward shows monotonicity of the Hawking mass on null surfaces along a flow defined by the condition 
that the expansions $\Theta^{+}$ and $\Theta^{-}$ are constant on the level sets \cite{Hay}. More precisely, for the 
surfaces spanned, say, by past directed null geodesics emanating from the future apparent horizon, it is the condition 
that $\Theta^{-}$ is constant on the level sets which defines the foliation while the constancy of $\Theta^{+}$ just 
defines a scaling of $\vec L^{+}$ convenient for the computation. Since an analogous monotonicity
statement holds along the other null congruence as well it seems that, by taking linear combination of the null vectors 
one can also obtain monotonicity along the corresponding spacelike directions. While this is not true in general (the 
statements on this issue in \cite{Hay} are not correct), it is nevertheless true in special cases, in particular when 
$\vec M$ is just the sum of the null vectors as in (\ref{lsum}). Therefore the analogy between Haywards result and 
the result of Malec, Mars and Simon in the spacelike case (as reformulated above) is not an incidence. It should be 
possible to write these results in a unified manner in a suitable formalism. \\
\\ {\Large\bf Acknowledgement}\\ \\
\noindent
W.S. was supported by Austrian FWF project Nr. P14621-Mat.


\begin{thebibliography}{99}

\setlength{\itemsep}{-0.6 ex}	

\bibitem{RP} R. Penrose,
       Ann. N.Y. Acad. Sci. {\bf 224}, 125 (1973).
\bibitem{HB} H. Bray, Jour. Diff. Geom. {\bf 59}, 177 (2001).
\bibitem{RG} R. Geroch,
              Ann. N.Y. Acad. Sci. {\bf 225}, 108 (1973).
\bibitem{JW} P.S. Jang and R.M. Wald,
               J. Math. Phys.(N.Y.) {\bf 18}, 41 (1977).
\bibitem{HI} G. Huisken, T. Ilmanen,
              Jour. Diff. Geom. {\bf 59}, 353 (2001).
\bibitem{MMS} E. Malec, M. Mars and W. Simon, Phys. Rev. Lett. {\bf 88}, 12102 (2002).
\bibitem{SH} S. Hawking,   J. Math. Phys. (N.Y.) {\bf 9}, 598 (1968).
\bibitem{LV} M. Ludvigsen, J. A. G. Vickers,
            J. Phys. A: Math. Gen. {\bf 16}, 3349 (1983).
\bibitem{GB} G. Bergqvist,
           Class. Quantum Grav. {\bf 14}, 2577 (1997).
\bibitem{Hay} S. Hayward,
             Class. Quantum Grav. {\bf 11}, 3037 (1994).

\end{thebibliography}
\end{document}